\documentclass[]{aastex631}
\usepackage{amsmath,amssymb}
\usepackage{graphicx}
\usepackage{color}

\def\coef{\rho_{\mathrm{Pearson}}}
\def\Lbol{\mathrm{L_{bol}}}
\def\MBH{\mathrm{M_{BH}}}
\def\REdd{\xi_\mathrm{Edd}}
\def\CGalDust{{\rm r-\left<W1\right>}}
\def\Cmean{{\left<C\right>}}
\def\CPlane{{$C_{\rm min}$--$\Delta C$}{\ }}

\begin{document}

\title{Long-term Mid-infrared Color Variations of Narrow-Line Seyfert 1 Galaxies}

\author[0009-0003-1518-6186]{Jiahua Wu}
\affiliation{Department of Astronomy, Guangzhou University, Guangzhou 510006, China} 
%  \email{jhwu@e.gzhu.edu.cn}

\author{Huifang Xie}
\affiliation{Department of Physics, Guangzhou University, Guangzhou 510006, China} 

 \author[0000-0002-4757-8622]{Liming Dou}
 \affiliation{Department of Astronomy, Guangzhou University, Guangzhou 510006, China}
 \correspondingauthor{Liming Dou}
 \email{doulm@gzhu.edu.cn}

 \author{Yanli Ai}
 \affiliation{College of Engineering Physics, Shenzhen Technology University, Shenzhen 518118, People’s Republic of China}

 \author{Tinggui Wang}
 \affiliation{Department of Astronomy, University of Science and Technology of China, Hefei, 230026, China}
 
 \author{Xinwen Shu}
 \affiliation{Department of Physics, Anhui Normal University, Wuhu, Anhui 241002, China}

 \author{Ning Jiang}
 \affiliation{Department of Astronomy, University of Science and Technology of China, Hefei, 230026, China}
 
\author[0000-0001-6947-5846]{Luis C. Ho}
\affiliation{Kavli Institute for Astronomy and Astrophysics, Peking University, Beijing 100871, China}
\affiliation{Department of Astronomy, School of Physics, Peking University, Beijing 100871, China}

\author{Junhui Fan}
\affiliation{Department of Astronomy, Guangzhou University, Guangzhou 510006, China} 

\begin{abstract}

We present a systematic investigation of long-term mid-infrared (MIR) color variability in 1,718 Narrow-Line Seyfert 1 galaxies (NLSy1s) using 14-year \textit{WISE}/NEOWISE monitoring data. Through Pearson correlation analysis between photometric magnitude and color, we identify: (1) a radio-quiet NLSy1 (RQ-NLSy1) population comprising 230 bluer-when-brighter (BWB) sources, 131 redder-when-brighter (RWB) sources, and 1,323 objects showing weak or statistically insignificant color variations; and (2) a radio-loud NLSy1 (RL-NLSy1) population containing 5 BWBs, 2 RWBs, and 27 sources with weak/no color variations. Our analysis reveals that the BWB tendency strengthens significantly in galaxies with redder mean MIR colors $\rm \left<W1-W2\right>$ and lower starlight contamination. Furthermore, this color-change pattern demonstrates that the most bolometric luminous sources exhibit the most pronounced BWB behavior. While similar trends exist for black hole mass and Eddington ratio, bolometric luminosity appears to be the primary physical driver. Potential origins of these variations (e.g., host galaxy contribution, accretion disk variability, and dust reprocessing) are discussed. We conclude that temperature-dependent dust reprocessing dominates the observed BWB, RWB, and no/weak variation patterns. This interpretation may also apply to similar MIR color variations observed in other extragalactic MIR transients, such as tidal disruption events, ambiguous nuclear transients, and changing-look AGNs. In addition, we find no significant difference in long-term MIR color variations between RL-NLSy1s and RQ-NLSy1s, however, RL-NLSy1s show significantly greater dispersion in intrinsic variability amplitude compared to RQ-NLSy1s due to jet-induced complexity, where non-thermal synchrotron emission from relativistic jets obscures thermal dust signatures.

\end{abstract}

\keywords{Infrared astronomy(786); Seyfert galaxies(1447); Black holes(162)}

\section{Introduction} \label{sec:intro}

The dominant infrared emission process in active galactic nuclei (AGN; \citealt{Seyfert1943,Schmidt1963}) involves thermal reprocessing of accretion disk UV/optical photons by circumnuclear dust \citep{Suganuma2006, Koshida2014, Mandal2018}, producing a spectral energy distribution (SED) peaking in the mid-infrared (MIR) band. The Wide-field Infrared Survey Explorer (\textit{WISE}; \citealt{Wright2010}) provides critical diagnostics through its MIR bands, where the color criterion $\rm W1-W2 \geq 0.8$ ($\rm W1=3.4~{\mu}m$, $\rm W2=4.6~{\mu}m$) reliably identifies AGNs via their hot dust continuum \citep{Assef2013}. MIR color variability serves as a unique probe \textbf{for} AGN physics. The primary mechanism involves accretion luminosity changes reverberating through the dust torus, thereby altering MIR colors \citep{Son2023}. However, significant contamination, such as the direct emission from the accretion disk (\citealt{Shakura1973}) and/or the host galaxy starlight (\citealt{Son2022, Fan2024}), should be also considered. 

Narrow-Line Seyfert 1 galaxies (NLSy1s) are a subclass of AGNs characterized by their narrow permitted emission lines, specifically full width at half maximum (FWHM) of H$\beta$ $\lesssim 2000$~km~s$^{-1}$, weak [O~\textsc{iii}] emission ([O~\textsc{iii}]/H$\beta$ $\lesssim 3$), and strong Fe~\textsc{ii} multiplet emission \citep{Osterbrock_and_Pogge1985,Goodrich1989,Veron2001}.
They predominantly hosted by spiral/disk galaxies undergoing rapid star-formation activity \citep[e.g.][]{Crenshaw2003,Jarvela2018,Varglund2022,Varglund2023}.
Their unique high-energy properties include larger soft X-ray ($\rm <2~keV$) variability amplitude, steeper X-ray spectra \citep[][]{Leighly1999,Grupe2004}, and smaller optical variability amplitude \citep{Ai2010,Rakshit2017}.
These observational signatures suggest the presence of supermassive black holes (SMBH) with relatively low-mass  ($10^6-10^7 {\rm M_{\odot}}$) accreting near or above the Eddington limit. Consequently, NLSy1s serve as critical laboratories for investigating the rapid SMBH growth mechanisms and possible early evolutionary stages of AGN activity \citep{Kawakatu2007}.

Recent studies have revealed two distinct long-term MIR color variation (MCV) patterns in NLSy1s, including both redder-when-brighter (RWB) and bluer-when-brighter (BWB) behaviors \citep{Caccianiga2015,Rakshit2019}. Interpretation of these variations requires careful disentanglement of different origins.
\citet{Rakshit2019} attribute RWB in NLSy1s to AGN radiation dominance (see also \citealt{Yang2018}). In contrast, based on a low-$z$ ($0.15 < z < 0.4$) MIR variability study, \citet{Son2022} find that lower luminosity AGNs exhibit RWB behavior and interpret this as evidence for dominance by blue starlight from the host galaxy.

Though most NLSy1s are radio-quiet (RQ-NLSy1s; \citealt{Zhou2002}), growing evidence reveals a population of radio-loud NLSy1s (RL-NLSy1s) with relativistic jets \citep[e.g.,][]{Yuan2008,Jiang2012}. $\gamma$-ray detections by \textit{Fermi}-Large Area Telescope in some RL-NLSy1s confirm the presence of relativistic jets \citep[e.g.,][]{Abdo2009,DAmmando2015,Paliya2018}, indicating significant non-thermal synchrotron contributions in MIR band\citep{Kozlowski2016_MIR}.
Jetted AGNs typically exhibit larger variability amplitudes \citep{Rakshit2017,Mao2021ApJS}, stronger short-term fluctuations \citep{Zhang2024Univ,Mao2021ApJS}, and leading to more complex MCV.
\citet{Anjum2020} found a predominantly BWB trend in MIR for $\gamma$-blazars. They interpreted this as being related to complex processes intrinsic to the jets, such as the injection of fresh electrons possessing a harder energy distribution than earlier cooled populations \citep{Kirk1998,Mastichiadis2002}. Jet-related processes, potentially also occurring in RL-NLSy1s (see also \citealt{Singh2025arXiv}), may contribute to the observed differences in MCV compared to RQ-NLSy1s. Thus, the drivers of MCV in NLSy1s remain unclear and warrant further investigation.

% MCVs  CL-AGNs, ANTs, TDEs
Tidal disruption event (TDE) occurs when a star approaches sufficiently close to a SMBH for tidal forces to tear it apart, resulting in a luminous nuclear flare. These events generate unique UV-to-optical flares with a sharp brightness rise and a luminosity evolution that follows a characteristic $t^{-5/3}$ power-law decay (\citealt{Gezari2021}). In systems containing circumnuclear dust, UV-optical emissions are absorbed and thermally re-emitted by the dust in the MIR band, producing detectable infrared echoes \citep[][]{Dou2016,Dou2017,Jiang2016,Velzen2016}. Another class of events, changing-look AGNs (CLAGNs), exhibit similar MIR variability but originate from accretion state transitions (see review in \citealt{Ricci2023}), and reliable distinction from TDEs typically requires UV/optical information \citep{Zabludoff2021}.
Additionally, there exists a class of nuclear outbursts that share characteristics of both TDEs and CLAGNs \citep[e.g.,][]{Blanchard2017,Petrushevska2023,Neustadt2020,Trakhtenbrot2019}. Due to the difficulty in determining their nature based solely on UV/optical data, these events are commonly classified as ambiguous nuclear transients (ANTs, \citealt{Blanchard2017}). They may represent TDEs occurring in AGN environments \citep[][]{Blanchard2017} or enhanced accretions onto SMBHs \citep[][]{Trakhtenbrot2019}. \cite{Yao2025} (hereafter Yao25) systematically analyzed a sample of MIR-bright nuclear transients, developing a robust classification scheme based on quantitative measurements of MIR color evolution rates. Comparative studies between these MIR-bright nuclear transients and the MCV of NLSy1s are particularly insightful, as both phenomena originate in dust-rich environments but through fundamentally different physical mechanisms.

In this work, we perform a comprehensive analysis of long-term MCV in a large sample of NLSy1s using multi-epoch \textit{WISE} data, and compare it with MIR outbursts from Yao25. We aim to constrain the dominant physical mechanisms driving the MCV in NLSy1s, and to explore their connection with black hole mass $\MBH$, Eddington ratio $\REdd$, and bolometric luminosity $\Lbol$.
In Section~\ref{sec:data} and Section~\ref{sec:classification}, we briefly describe the sample selection and details of the mid-infrared lightcurve data analysis, while the properties for MCV are elaborated in Section~\ref{sec:properties}.
We discussed and \textbf{concluded} the properties in Section~\ref{sec:discussion} and Section~\ref{sec:conclusions}, respectively.
Throughout, we assume a standard flat $\Lambda\rm{CDM}$ cosmology with $H_0=70~\rm{km}~\rm{s}^{-1}~\rm{Mpc}^{-1}$, $\Omega=0.3$, and $\Omega_{\Lambda}=0.73$; all magnitudes are in the Vega system.

\section{Sample Selection And Data Reduction} \label{sec:data}

We used a parent sample of 22,656 spectroscopically confirmed NLSy1s from the \citet{Paliya2024} catalog, which is based on SDSS DR17. To mitigate potential biases from cosmological evolution and uncertainties in $K$-correction, we made a redshift cut of $z < 0.3$. The resulting sample was cross-matched with `AllWISE' and `NEOWISE-R' light curves using a 3\arcsec\ positional tolerance, yielding 4,032 unique counterparts. For each source, we extracted 20--25 epochs of W1 and W2 photometry. Within each epoch comprising 10--20 exposures, we applied stringent quality criteria:
\begin{enumerate}
\item Only detections from framesets with \texttt{qual\_frame}$>$0 were retained. Framesets with \texttt{qual\_frame}=0 often indicate issues such as spurious detections of noise, transient events, or scattered light,

\item The number of point spread function (PSF) components used in profile fit for a source must be less than 3 (\texttt{nb} $\leq$ 2),

\item The best quality single-exposure image ($\texttt{qi\_fact}$ = 1), and frames are unaffected by known artifacts (\texttt{cc\_flags} = `00'), are not potentially contaminated by scattered moonlight (\texttt{moon\_masked} = 0) and are not actively de-blended (\texttt{na} = 0),

\item We selected database entries with $\texttt{saa\_sep}>0$ to avoid framesets taken when the WISE spacecraft was within the boundaries of the boundary of the South Atlantic Anomaly (SAA).
\end{enumerate}
We performed 5-$\sigma$ clipping (where $\sigma$ represents the standard deviation of the light curve) within each epoch to remove photometric outliers. Measurements passing these criteria were temporally averaged to derive epoch-specific magnitudes. Additionally, we required both a minimum of five photometric measurements per observing epoch and, after applying this filter, at least five remaining epochs per source.

The intrinsic variability amplitude $\sigma_m$ was quantified following the methodology of \citet{Sesar2007}. We retained only sources exhibiting $\sigma_m > 2\epsilon$ (where $\epsilon$ is the measurement uncertainty) in both W1 and W2 bands. This selection yielded well-characterized long-term light curves for 1,718 NLSy1s, spanning the full \textit{WISE}/NEOWISE mission duration from January 2010 to August 2024. % (14-year baseline).

\begin{figure}[hbt!]
\centering
% \plotone{wiselc_demos.pdf}
 \includegraphics[width=0.7\linewidth]{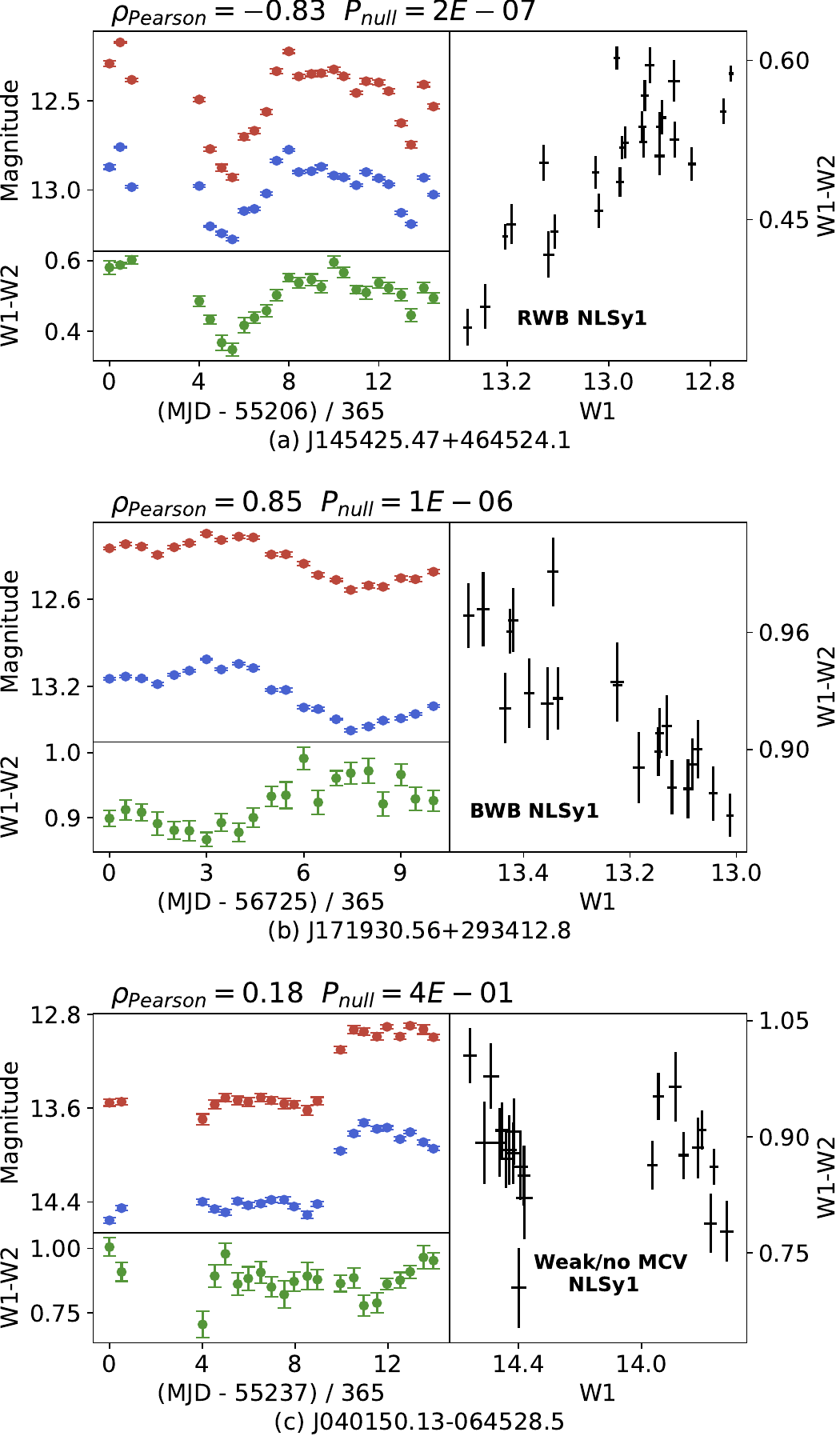}
\caption{Example MIR light curves for each classification in Section~\ref{sec:classification}. For each panel, light curves (Upper left, Blue in W1 band and Red in W2 band), color variations (Lower left), and color variations following the photometric magnitudes in W1 band (Right) are shown.
\label{fig:lcs}}
\end{figure}

In order to characterise the MCV behavior of an individual NLSy1, we constructed MIR color variation curves (defined as $C = {\rm W1-W2}$) for each source and performed Pearson correlation analysis between \textbf{$C$} and W1-band photometric magnitude.
This yielded the Pearson correlation coefficient $\coef$ and the corresponding null-hypothesis probability $P_\text{null}$ for each source (Figure~\ref{fig:lcs}). If $C$ decreases (i.e. bluer in color) with increasing brightness in the light curve (i.e. BWB behavior), $\coef$ is predicted to be positive. However, if $C$ becomes redder as brightness increases (i.e. RWB behavior), $\coef$ is predicted to be negative.
We employ $\coef$ to quantify the degree of MCV in NLSy1s. A value approaching +1 indicates a more pronounced BWB trend, while a value approaching -1 signifies a stronger RWB trend. Thus, $\coef = +1$ and $\coef = -1$ correspond to ``extreme BWB'' and ``extreme RWB'', respectively.
For the 1,718 NLSy1s in our sample, the average of $\coef$ is 0.06 with a sample standard deviation is 0.44, and the maximum and minimum values are 0.98 and -0.92, respectively. This indicates that $\coef$ centered near zero with tails extending toward both BWB and RWB extremes.
To assess the reliability of $\coef$, we accounted for photometric uncertainties by propagating them into the estimate of statistical error for each source. The resulting median 1-$\sigma$ uncertainty of $\coef$ across the sample is 0.19. This implies that sources with $|\coef| > 0.6$ can be considered statistically significant for exhibiting specific MCV behavior (BWB or RWB).

To investigate the relation between mid-infrared color \textbf{variation} MCV and AGN physical parameters, we obtained the following parameters for our sample: black hole mass ($\MBH$), bolometric luminosity ($\Lbol$), Eddington ratio ($\REdd$), and radio-loudness (RL) from \citet{Paliya2024}.

\section{Classification of Mid-infrared Color Variation} \label{sec:classification}

Using the radio-loudness parameter RL, we classify sources as RQ-NLSy1s ($\text{RL} < 10$) or RL-NLSy1s ($\text{RL} \geq 10$), where the RL is defined as the ratio of the rest-frame flux densities at 5~GHz and 4400~$\AA$ \citep{Paliya2024}. The 5~GHz flux was derived by extrapolating the 1.4~GHz flux from the FIRST survey \citep{White1997} using a spectral index of $\alpha=0.5$. Based on the Pearson correlation coefficient $\coef$, corresponding null-hypothesis probability $P_\text{null}$, we define three distinct MCV behaviors:
\begin{enumerate}
\item[(1)] \textbf{BWB NLSy1s}: 235 sources (230 RQ-NLSy1s and 5 RL-NLSy1s) exhibiting strong positive correlation ($\coef > 0.6$, $P_\text{null}<0.05$);
\item[(2)] \textbf{RWB NLSy1s}: 133 sources (131 RQ-NLSy1s and 2 RL-NLSy1s) exhibiting strong negative correlation ($\coef < -0.6$, $P_\text{null} < 0.05$);
\item[(3)] \textbf{Weak/no MCV NLSy1s}: 1,350 sources (1,323 RQ-NLSy1s and 27 RL-NLSy1s) showing either weak ($|\coef| \leq 0.6$) or statistically insignificant ($P_\text{null} \geq 0.05$) MCV.
\end{enumerate}

Figure~\ref{fig:lcs} presents three representative examples of MIR light curves: J145425.47+464524.1 (RWB NLSy1), J171930.56+293412.8 (BWB NLSy1), and J040150.13-064528.5 (Weak/no MCV NLSy1).

\section{Properties of NLSy1s with Mid-Infrared Color Variation} \label{sec:properties}

\subsection{Intrinsic variability amplitude, colors, and AGN parameters}\label{sec:relations}

Figure~\ref{fig:mcv_iav} presents the correlation coefficient ($\coef$) versus intrinsic variability amplitude $\sigma_m$ in the W1 and W2 bands. A clear trend is observed in the RQ-NLSy1s: as $\coef$ increases from -1 (extreme RWB) to +1 (extreme BWB), the $\sigma_m$ in the W2 band  systematically decreases relative to that in W1.  The Pearson correlation coefficients between $\coef$ and $\log~\sigma_m$ are 0.08 (p-value=0.001) for W1 and -0.46 (p-value=$4\times10^{-88}$) for W2 in RQ-NLSy1s.

RL-NLSy1s display similar MCVs distribution patterns to their radio-quiet counterparts, with two notable distinctions.
First, RL-NLSy1s exhibiting weak/no MCV ($|\coef| < 0.6$) show markedly broader intrinsic variability amplitude $\sigma_m$ distributions. The dispersion in $\sigma_m$ values, as measured by the sample standard deviation, is approximately twice as large in RL-NLSy1s compared to RQ-NLSy1s in both W1 (0.103 vs. 0.057) and W2 (0.103 vs. 0.062) bands. This increased dispersion is statistically confirmed by Levene's test for equality of variances (p-value=$3.1 \times 10^{-4}$ for W1; p-value=$4.9 \times 10^{-3}$ for W2).
Second, this subsample includes several RL-NLSy1s with extreme $\sigma_m$ values that exceed those observed in RQ-NLSy1s by factors of 2-3. Given that radio-loud AGNs are well-known to exhibit substantial non-thermal, jet-driven continuum variability \citep{Kozlowski2016_MIR}, the enhanced dispersion in  intrinsic variability amplitude for weakly color variable RL-NLSy1s likely reflects the superposition of jet-dominated continuum emission with thermal dust reprocessing signals.

\begin{figure}[hbt!]
\centering
%\plotone{rpvsvar.pdf}
 \includegraphics[width=0.6\linewidth]{rpvsvar.pdf}
\caption{Distributions of the MCV Pearson correlation coefficient $\coef$ versus intrinsic amplitude of variability $\sigma_m$ in W1 band (Top) and W2 band (Bottom). Red rhombuses represent RL-NLSy1s, while black dots denote RQ-NLSy1s. Dashed lines are marked at $\coef = \pm0.6$. \label{fig:mcv_iav}}
\end{figure}

\begin{figure}[hbt!]
\centering
%\plotone{color_vs_rho.pdf}
 \includegraphics[width=0.6\linewidth]{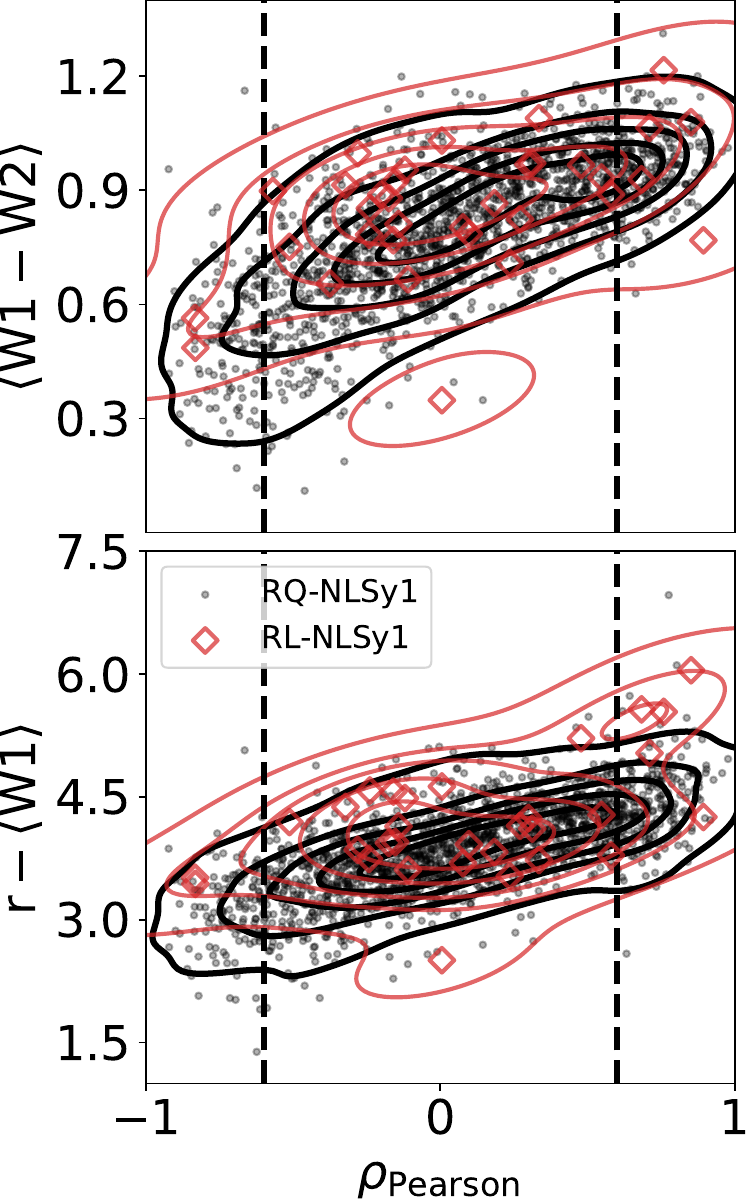}
\caption{
Distributions of the MCV Pearson correlation coefficient $\coef$ versus mean MIR color $\rm \left<W1-W2\right>$ (Top) and $\CGalDust$ (Bottom).  Symbols are consistent with those in Figure~\ref{fig:mcv_iav}. 
\label{fig:color_vs_rho}
}
\end{figure}

In contrast to the stronger correlation observed between $\sigma_m$ and $\coef$ in W2 band, a weaker correlation is evident in the W1 band. This difference may arise because the W1-band flux is more susceptible to contamination from both the accretion disk and the host galaxy. We therefore investigate how $\coef$ relates to the color $\rm \langle W1-W2\rangle$ (tracing dust temperature) and $\CGalDust$ (tracing relative dust contribution).
Figure~\ref{fig:color_vs_rho} demonstrates that for RQ-NLSy1s, $\coef$ exhibits strong positive correlations with both $\rm \left<W1-W2\right>$ (r=0.72, p-value=$1.3\times10^{-270}$) and $\CGalDust$ (r=0.69, p-value=$8.5\times10^{-233}$). The distributions for RL-NLSy1s show no significant deviation from those of RQ-NLSy1s. This indicates that MCV in our NLSy1 sample is linked to changes in the relative contributions of dust, host galaxy, and accretion disk, as discussed in detail in Section~\ref{sec:discussion}.

\begin{figure*}[hbt!]
\centering
\includegraphics[width=1.0\textwidth]{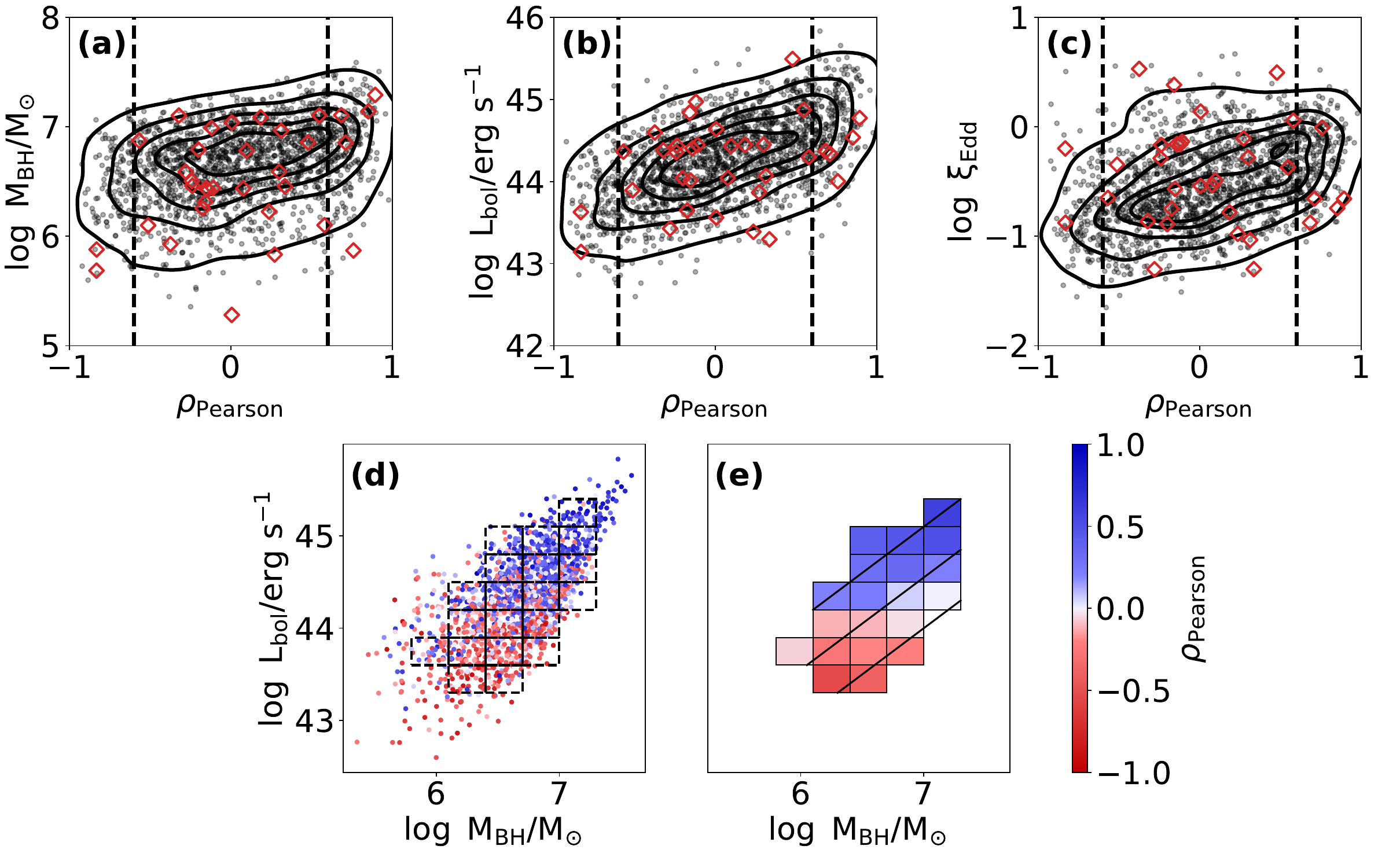}
\caption{MCV Pearson correlation coefficients v.s. AGN parameters in NLSy1s. Panel (a)--(c): $\coef$ v.s. black hole mass $\MBH$, bolometric luminosity $\Lbol$, and Eddington ratio $\REdd$. 
RQ-NLSy1s and RL-NLSy1s are shown as black dots and red rhombuses, respectively. Dashed lines indicate $\coef=\pm 0.6$ for reference.
Panel (d): Distribution of individual RQ-NLSy1s in the $\MBH$-$\Lbol$ space, color-coded by $\coef$ values. Dashed rectangles denote 0.3~dex $\times$ 0.3~dex bins ($\geq$20 sources each). Panel (e): Median $\coef$ within each bin, color-coded to show the dominant variability trend. Diagonal lines indicate constant Eddington ratio $\REdd$, with values increasing from bottom to top.
\label{fig:rho_agn_properties}}
\end{figure*}

We investigate the relationship between MCV and fundamental AGN properties (black hole mas $\MBH$, Eddington ratio $\REdd$, and Bolometric luminosity $\Lbol$) taken from \citet{Paliya2024}.
As shown in Figure~\ref{fig:rho_agn_properties} (panel a--c), the coefficient $\coef$ demonstrates significant positive correlations with black hole mass $\MBH$ (r=0.33, p-value=$8\times10^{-45}$), Eddington ratio $\REdd$ (r=0.42, p-value=$1\times10^{-73}$), and $\Lbol$ (r=0.56, p-value=$1\times10^{-141}$).
However, these parameters are intrinsically coupled ($\REdd \propto \Lbol/\MBH$).
To disentangle the effects of these intrinsically coupled parameters, we analyze the $\MBH$-$\Lbol$ plane for RQ-NLSy1s (Figure~\ref{fig:rho_agn_properties}, panels d–e). Binning the sample reveals that: (1), when controlling for $\Lbol$, $\coef$ shows no significant correlation with $\MBH$; (2), higher $\Lbol$ systematically increases $\coef$ values; (3), Diagonals of constant log($\Lbol$/$\MBH$) exhibit progressively stronger BWB trends, confirming that $\REdd$-driven variations are ultimately luminosity dependent. These results establish bolometric luminosity as the dominant physical driver of MCV, with dust temperature evolution shaping the observed BWB/RWB dichotomy.

\subsection{MIR Color Variation Distribution}

\begin{figure*}[hbt!]
\plotone{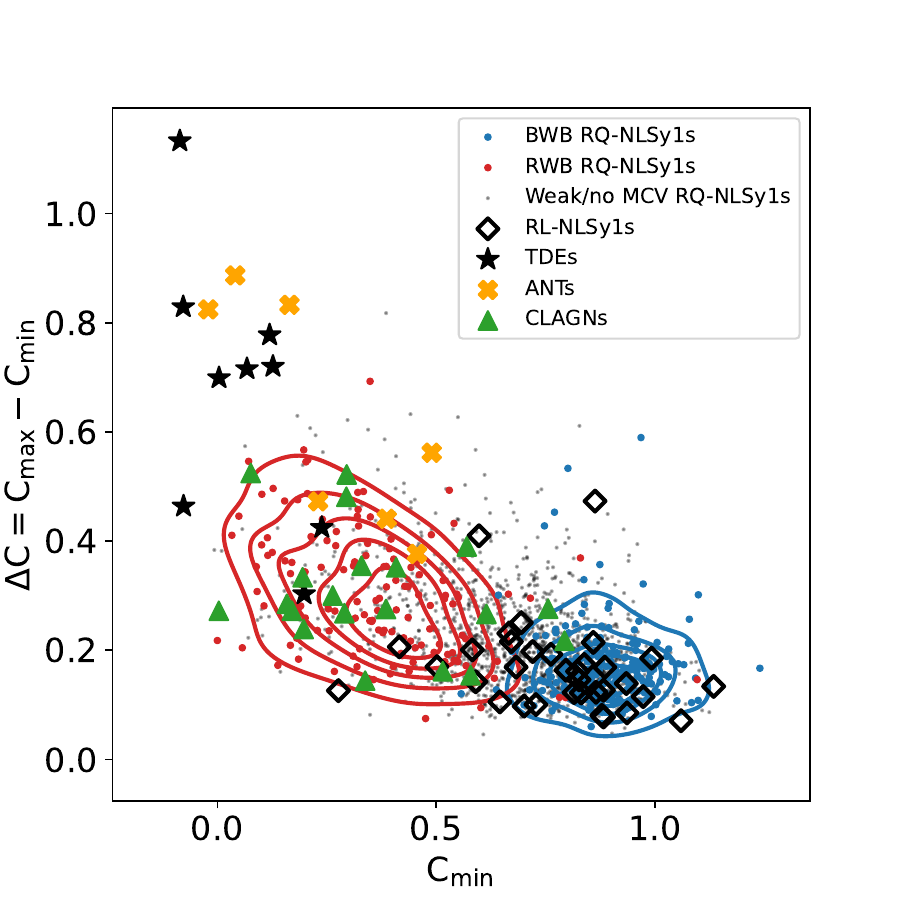}
\caption{
Distribution of the minimal MIR color ($C_{\rm min}$) versus maximal MIR color variation ($\Delta C=C_{\rm max}-C_{\rm min}$) for NLSy1s (red: RWB RQ-NLSy1s; blue: BWB RQ-NLSy1s; gray: weak/no MCV RQ-NLSy1s). Overplotted MIR outbursts from Yao25 include CLAGNs (green triangles), TDEs (black pentagrams), and ANTs (yellow crosses).
\label{fig:cplane}}
\end{figure*}

Figure~\ref{fig:cplane} presents the distribution of the minimal MIR color $C_{\rm min}$ versus the maximal MIR color variation $\Delta C = C_{\rm max} - C_{\rm min}$ for NLSy1s.
For each source, $C_{\rm min}$ ($C_{\rm max}$) is defined as the minimum (maximum) MIR color measured over the monitoring period, corresponding to the bluest (reddest) observed state.
The RWB RQ-NLSy1s exhibit a broad distribution centered at $C_{\rm min} = 0.36$ and $\Delta C = 0.29$, while the BWB RQ-NLSy1s form a tight cluster in the lower-right quadrant with median values of $C_{\rm min} = 0.88$ and $\Delta C = 0.15$. This demonstrates that BWB RQ-NLSy1s consistently display redder MIR colors ($C_{\rm min} \gtrsim 0.7$), as shown in the top panel of Figure~\ref{fig:color_vs_rho}, whereas RWB RQ-NLSy1s exhibit both larger and more widely distributed $\Delta C$ values. Weakly variable RQ-NLSy1s occupy an intermediate region, predominantly near the BWB population. RL-NLSy1s (black diamonds) show a similar distribution but align more closely with the BWB RQ-NLSy1 group.

For comparative analysis with nuclear transient phenomena, we incorporate MIR outbursts from Yao25's catalog (Figure~\ref{fig:cplane}), including TDEs, CLAGNs, and ANTs. Applying identical processing as in Section~\ref{sec:data}, we analyze 36 MIR outbursts with redshifts $<0.12$. TDEs occupy the upper-left quadrant, characterized by extremely blue color states ($C_{\rm min} < 0.24$) and large variations ($0.3 < \Delta C < 1.1$, the median of $\Delta C$=0.7). ANTs exhibit similarly high $\Delta C$ values but span a broader range in $C_{\rm min}$. CLAGNs predominantly populate intermediate regions, showing substantial overlap with RWB RQ-NLSy1s.

\section{Discussion} \label{sec:discussion}
% AGN disk emission, host galaxy
\subsection{MCV in NLSy1s}

The MIR emission in AGNs is predominantly powered by thermal reprocessing of UV-optical radiation from the accretion disk by surrounding hot dust. Although dust emission dominates the MIR regime, contributions from the accretion disk and host galaxy can modulate observed MCV.

Our sample of NLSy1s reveals diverse MCV patterns: while most sources exhibit weak or no MCV, distinct subsets display pronounced BWB or RWB behavior. This diversity suggests that multiple emission components: hot dust, accretion disk, and host galaxy, collectively obscure MCV signatures. Additionally, the differing response times of the W1 and W2 bands to continuum variations from the accretion disk further influence MCV characteristics\citep{Son2022}. 
Standard AGN SEDs, characterized by the big blue bump and infrared bump, indicate that accretion disk emission contributes to the W1 band only through its red tail \citep[e.g.,][]{Ezhikode2017}. If disk variability were the primary driver of MIR fluctuations, we would expect correspondingly strong UV-optical variations. However, NLSy1s lack systematically strong optical variability \citep{Ai2010,Rakshit2017}. We therefore treat direct contributions from both the accretion disk and host galaxy as approximately invariant across the MIR band.

AGNs exhibit characteristically red MIR color (W1-W2 $>$ 0.8; \citealt{Assef2013,Stern2012}). As shown in the top panel Figure~\ref{fig:color_vs_rho}, RQ-NLSy1s with strong BWB behavior typically display larger mean MIR color values ($\Cmean_{median}$=0.96), while those with strong RWB behavior are systematically bluer ($\Cmean_{median}$=0.52). This color segregation aligns with the positive correlations between $\coef$ and AGN physical parameters (Figure~\ref{fig:rho_agn_properties}), particularly the fundamental dependence on bolometric luminosity ($\Lbol$; see Section~\ref{sec:relations} and Figure~\ref{fig:rho_agn_properties}). When the variable nuclear component dominates (i.e., at high $\Lbol$), NLSy1s predominantly exhibit BWB behavior.

The correlation between $\coef$ and $\CGalDust$ (Figure~\ref{fig:color_vs_rho}, bottom panel) directly probes the relative contributions of AGN dust emission versus accretion disk and host galaxy emission. BWB RQ-NLSy1s show systematically larger $\CGalDust$ values, indicating a higher fraction of hot dust emission. In contrast, RWB RQ-NLSy1s exhibit smaller $\CGalDust$, suggesting non-negligible contributions from the host galaxy or accretion disk. Collectively, these observations imply that BWB behavior is intrinsic to dust emission variability, while dilution by host galaxy or disk emission weakens the observed MCV signature.

We interpret these observational patterns through the framework of dust temperature evolution.
RWB sources typically host lower-mass black holes with modest accretion rates, sustaining cooler dust temperatures. When accretion surges occur, the dust is heated to intermediate temperatures, boosting long-wavelength emission relative to the shorter-wavelength baseline.
Conversely, BWB sources harbor higher-mass black holes with more vigorous accretion, maintaining hotter dust temperatures. Further heating shifts the SED peaks blueward, amplifying short-wavelength flux relative to the redder baseline.

To model these effects, we employed a blackbody spectrum with temperature $T_d$ ranging from 200 to 2000~K for the circumnuclear hot dust component, while accounting for constant MIR flux contributions from the accretion disk or host galaxy.
The total MIR flux at specific wavelength $\lambda$ is given by 
\begin{equation}
F_{\lambda}(T_d | A_{\rm bb}, F_{\lambda,\mathrm{invar}})
= F_{\lambda,\mathrm{invar}} + A_{\rm bb} \cdot \frac{2hc}{\lambda^3} \cdot \frac{1}{\mathrm{exp}(\frac{hc}{\lambda k T_d}) - 1}
\end{equation}
where $h$ is the Planck constant, $c$ is the speed of light, $k$ is the Boltzmann constant, and $T_d$ is the dust temperature. The parameter $A_{\rm bb}$ is a dimensionless factor of blackbody radiation. $F_{\lambda,\mathrm{invar}}$ represents the wavelength-dependent invariant flux from the accretion disk or host galaxy.
To simulate varying levels of contamination from the accretion disk or host galaxy, we adopted a fixed normalization $A_{\rm bb}$ and set different values for the invariant flux $F_{\lambda,{\rm invar}}$. To compare our model directly with the observations, we converted the modeled fluxes into Vega magnitudes using the WISE zeropoints (306.681~Jy for W1 and 170.663~Jy for W2, \citealt{Wright2010}).
Based on the result of \cite{Son2022}, the fractional contribution of starlight in the W1 band is widely distributed, with an average of 22\% (see Figure~9 of \citealt{Son2022}). We set $F_{\mathrm{W1,invar}}$ to 0.093~mJy and 0.465~mJy, corresponding to 10\% and 50\% starlight fractions for a source with total W1 flux of 0.93~mJy (i.e., W1 = 13.7 mag, the sample median). The normalization $A_{\mathrm{bb}} = 7\times10^{-20}$ was chosen to produce W1 magnitudes covering the sample median.
% We explored two scenarios: one in which the observed MIR flux has a substantial contribution from starlight of the host galaxy ($F_{\mathrm{W1, invar}} = 0.8$~mJy), and another in which the circumnuclear hot dust emission dominates ($F_{\mathrm{W1, invar}} = 0.05$~mJy).
The corresponding $F_{\mathrm{W2, invar}}$ was determined from the relation $F_{\mathrm{W1, invar}}/F_{\mathrm{W2, invar}} = 1.67$, a ratio derived from the spiral galaxy template of \citet{Assef2010}. In the presented scenarios, we simplistically attribute the invariant flux solely to host starlight for illustrative purposes. In practice, both $A_{\mathrm{bb}}$ and the invariant component can be adjusted to match the actual light curves of individual sources.

\begin{figure}[hbt!]
\plotone{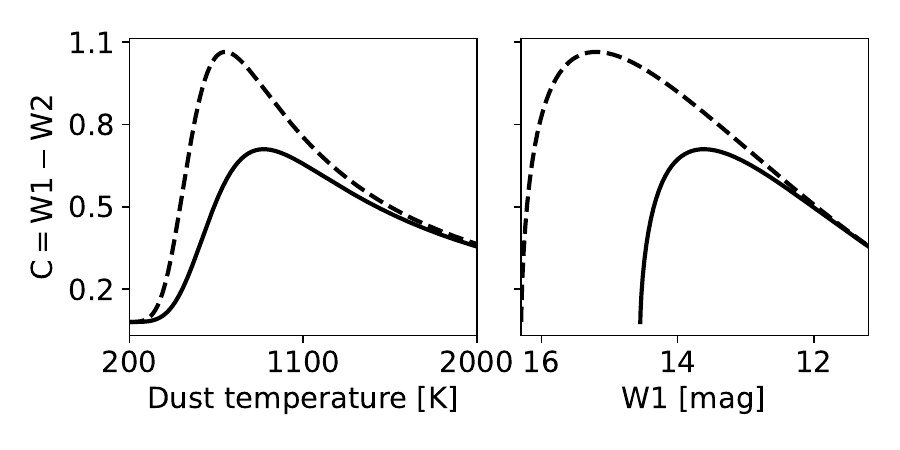}
\caption{
Left: Modeled MIR color, $C=\mathrm{W1-W2}$, as a function of dust temperature for circumnuclear hot dust, assuming blackbody radiation. The parameter $A_{\rm bb}$ is fixed at $7\times10^{-20}$. The dashed ($F_{\rm W1,invar} = 0.093$~mJy) and solid ($F_{\rm W1,invar} = 0.465$~mJy) lines represent different levels of host galaxy emission. The corresponding $F_{\mathrm{W2, invar}}$ was determined from $F_{\mathrm{W1, invar}}/F_{\mathrm{W2, invar}} = 1.67$.
Right: Modeled color-magnitude diagram under the same assumptions as the left panel. The MIR color $C$ is shown as a function of W1-band magnitude.
\label{fig:model}}
\end{figure}

The results, presented in Figure~\ref{fig:model}, confirm that this simplified model successfully reproduces both RWB and BWB behaviors.
Notably, the model predicts larger color changes ($\Delta C$) for RWB sources at equivalent temperature variations—consistent with the observational trend shown in Figure~\ref{fig:cplane}, where RWB RQ-NLSy1s exhibit generally larger $\Delta C$ than their BWB counterparts.
Given that our analysis indicates the MIR emission in BWB sources is dominated by circumnuclear hot dust, we estimated their characteristic dust temperatures by fitting the average W1 and W2 flux densities with a single-temperature blackbody model. The resulting temperature distribution has a mean of 1107$\pm$84~K and spans a range from approximately 876~K to 1396~K.

% \subsection{Radio-loud NLSy1}

The MIR emission of RL-NLSy1s is significantly contaminated by non-thermal synchrotron radiation from relativistic jets \citep{Kozlowski2016_MIR}. Despite statistical limitations due to their smaller sample size, RL-NLSy1s exhibit overall trends consistent with RQ-NLSy1s (Figure~\ref{fig:mcv_iav}-\ref{fig:cplane}). However, RL-NLSy1s with large variability amplitude $\sigma_m$ typically display weak or no MCV (Figure~\ref{fig:mcv_iav}), suggesting complex emission mechanisms. Furthermore, their distribution in the \CPlane plane (Figure~\ref{fig:cplane}) broadly aligns with BWB RQ-NLSy1s, though with significant scatter.

While jetted AGNs (e.g., BL Lacertae objects and flat-spectrum radio quasars) often demonstrate BWB trends \citep{Anjum2020}, RL-NLSy1s may inherit similar behavior through jet-related processes  \citep[e.g.,][]{Kirk1998,Mastichiadis2002,Villata2004,Papadakis2007}. The observed weak MCV is likely due to the superposition of multiple emission mechanisms, which dilutes thermal dust reprocessing signatures.

\subsection{Comparing with MIR-bright optical outbursts from Yao25}

TDEs induce flare states in quiescent galaxies, triggering an infrared (IR) echo as the circumnuclear dust responds to the sudden outburst. Prior to the flare, the host galaxy typically dominates the observed radiation. As shown in Figure~\ref{fig:cplane}, the \CPlane distribution of TDEs extends the trend observed in RWB RQ-NLSy1s, supporting the interpretation that MCV in both classes is primarily driven by dust temperature evolution. TDEs exhibit significantly larger color changes $\rm \Delta C$ and reach extremely blue states, consistent with thermal emission from stellar debris accretion. The enhanced $\rm \Delta C$ arises from substantial dust temperature variations induced by the nuclear outburst. ANTs, despite their heterogeneity, largely occupy the same region in the \CPlane as TDEs. This overlap suggests that some ANTs may represent unclassified TDEs or unusual AGN flares.

CLAGNs, which transition spectroscopically between broad-line and narrow-line states accompanied by luminous continuum flares, populate an intermediate region in the \CPlane, exhibiting a distribution similar to RWB RQ-NLSy1s. While the mechanism driving their spectral transformations remains unclear, the presence of a pre-existing dusty torus implies that their color changes can similarly be attributed to changes in the dust temperature. This is driven by the delayed response of the dust to the AGN's variable radiation: the dust is heated by the continuum flare. Consistent with the temperature evolution framework, CLAGNs systematically display smaller $\Delta C$ compared to TDEs/ANTs due to their smaller $\delta W2$ (Figure 3 in \citealt{Yao2025}).

\section{Conclusions} \label{sec:conclusions}

We present a systematic analysis of MCV in 1,718 NLSy1s selected from \citet{Paliya2024}, utilizing 13--14 years multi-epoch \textit{WISE} monitoring. Our principal findings are as follows:
 
\begin{enumerate}

\item[(1)] MCV is ubiquitously driven by thermal dust reprocessing, with color variations tracing temperature changes in circumnuclear dust responding to accretion activity. Contamination from the accretion disk and host galaxy components yields two distinct behaviors: bluer-when-brighter trends in higher-temperature dust environments, and red-when-brighter trends in lower-temperature dust environments.

\item[(2)] The color variability coefficient $\coef$ strongly correlates with bolometric luminosity $\Lbol$, independent of black hole mass $\MBH$, confirming that higher accretion power preferentially induces BWB behavior.

\item[(3)] Transients and CLAGNs follow similar thermal reprocessing mechanisms. TDEs/ANTs exhibit larger color changes $\Delta C$ value due to extreme luminosity variations, while CLAGNs show intermediate behavior consistent with torus temperature evolution.

\item[(4)] RL-NLSy1s exhibit more dispersion in their intrinsic variability amplitude $\sigma_m$ compared to RQ-NLSy1s, attributed to non-thermal synchrotron emission from relativistic jets obscuring thermal dust signatures.

\end{enumerate}

These results establish dust temperature evolution as the primary driver of MCV in RQ-NLSy1s, with accretion parameters (particularly bolometric luminosity) dictating observed trends. The distinct behaviors between BWB/RWB systems provide a robust diagnostic for probing circumnuclear dust properties in AGN.

\begin{acknowledgments}
We are grateful to the anonymous referee for comments that have improved the quality of this paper.
This work was supported by the Joint Research Foundation in Astronomy under the cooperative agreement between the National Science Foundation of China and the CAS (U1731104 and U1731109), National Science Foundation of China (NSFC-12573110, 12433004, 12133001 and 11833007). LCH was supported by the National Science Foundation of China (12233001) and the China Manned Space Program (CMS-CSST-2025-A09). YLA was supported by the Natural Science Foundation of Top Talent of SZTU (GDRC202208) and the Shenzhen Science and Technology program (JCYJ20230807113910021). LMD also acknowledges the support from the Key Laboratory for Astronomical Observation and Technology of Guangzhou, the Astronomy Science and Technology Research Laboratory of the education department of Guangdong Province.
This research makes use of data products from the {\it Wide-field Infrared Survey Explorer (WISE)} and the {\it Near-Earth Object Wide-field Infrared Survey Explorer (NEOWISE)}. {\it WISE} is a joint project of the University of California, Los Angeles, and the Jet Propulsion Laboratory/California Institute of Technology; {\it NEOWISE} is a project of the Jet Propulsion Laboratory/California Institute of Technology. {\it WISE} and {\it NEOWISE} are funded by the National Aeronautics and Space Administration.

\end{acknowledgments}

\bibliography{ms_v2}{}
\bibliographystyle{aasjournal}
\end{document}